\documentstyle[epsf,emulateapj]{article}
\begin{document}

\leftline{\epsfbox{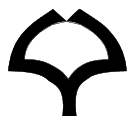}}
\vspace{-10.0mm}
\thispagestyle{empty}
{\baselineskip-4pt
\font\yitp=cmmib10 scaled\magstep2
\font\elevenmib=cmmib10 scaled\magstep1  \skewchar\elevenmib='177
\leftline{\baselineskip20pt
\hspace{12mm} 
\vbox to0pt
   { {\yitp\hbox{Osaka \hspace{1.5mm} University} }
     {\large\sl\hbox{{Theoretical Astrophysics}} }\vss}}

%
%
{\baselineskip0pt
\rightline{\large\baselineskip14pt\rm\vbox 
        to20pt{\hbox{OU-TAP-110}
               \hbox{astro-ph/0002010}           
\vss}}
}
\vskip15mm

\title {RADIATION SPECTRA FROM \\ADVECTION-DOMINATED 
ACCRETION FLOWS  \\IN A GLOBAL MAGNETIC FIELD}
\author { MOTOKI KINO$^{1}$, OSAMU KABURAKI and NAOHIRO YAMAZAKI}
\affil { Astronomical Institute, Graduate School of Science, 
Tohoku University, \\Aoba-ku, Sendai 980-8578, Japan;\\
kino@astr.tohoku.ac.jp, okabu@astr.tohoku.ac.jp, 
ietty@astr.tohoku.ac.jp}

\begin{abstract}
We calculate the radiation spectra from advection-dominated 
accretion flows (ADAFs), taking into account the effects of 
a global magnetic field.
Calculation is based on the analytic model for magnetized 
ADAFs proposed by Kaburaki, where a large-scale magnetic 
field controls the accretion process.
Adjusting a few parameters, we find that our model can well 
reproduce the observed spectrum of Sagittarius A$^{*}$. 
The result is discussed in comparison with those of  well-known 
ADAF models, where the turbulent viscosity controls the accretion 
process. 
\end{abstract}


\section{INTRODUCTION}
\footnotetext[1]{Present adress: Department of Earth and Space Science,
Graduate School of Science, Osaka University, Toyonaka, Osaka
560-0043, Japan\\
kino@vega.ess.sci.osaka-u.ac.jp}
The optically thin, advection-dominated accretion flows (ADAFs) have
been studied by a number of authors during past several years
(e.g. Narayan $\&$ Yi 1994, 1995a, b; Abramowicz et al.\ 1995;
Nakamura et al.\ 1997, Manmoto, Mineshige \& Kusunose 1997; Narayan et
al.\ 1998). 
 These models are very successful in
describing both spectra and dynamics of accreting black hole systems
such as those in binaries and in low-luminosity active galactic nuclei
(AGNs).  The observed spectra can be explained as follows.  The radio
emission is due to the synchrotron emission in turbulent magnetic
fields in the accretion flow.  These synchrotron photons serve as seed
photons for the inverse Compton process by hot electrons.
Once-scattered Compton photons are mainly distributed in the optical
band and twice-scattered Compton photons, in soft X-ray band.
Bremsstrahlung due to electron-electron and electron-proton collisions
gives rise to the observed hard X-ray spectra.

Thus, these ADAF models provide a good framework for understanding 
the observed spectra. 
In these models,  both angular momentum transfer and energy 
dissipation in the accretion flow is assumed to be undertaken by 
the turbulent viscosity whose size is specified by so-called  
$\alpha$ parameter. 
For this reason, hereafter we call this type of models the 
``viscous'' ADAF model in this paper. 
The magnetic fields are regarded as of turbulence origin and are 
described by another parameter $\beta$ which specifies the ratio 
of the magnetic pressure to the gas pressure.

However there is no reason to believe that the turbulent viscosity
is the only candidate that controls the accretion processes. 
Rather, it is quite natural to think that some types of global 
magnetic fields may play an essential role. 
Indeed, there are some evidences for the presence of such an 
ordered magnetic field in the central region of our Galaxy 
(e.g., Yusef-Zadeh, Morris \& Chance 1984). 
As Kato, Fukue \& Mineshige (1998) has pointed out the hydromagnetic 
turbulence in accretion disks may also generate global magnetic 
fields by dynamo processes due to the presence of helical motions.
In view of such circumstances, another type of ADAF model has been 
proposed by one of the present authors (Kaburaki 1999, 2000; hereafter 
referred to as K99 and K00).
In order to distinguish it from the above viscous ADAF models, 
hereafter we call it the ``resistive''  ADAF models since energy 
dissipation in the accretion flow is due to the electric resistivity 
and angular momentum transfer is supported not by the viscosity 
but by the magnetic stress of a large scale magnetic field.

The purpose of the present study is to calculate the expected 
radiation spectra from ADAFs in a global magnetic field 
based on the resistive ADAF model, in order to compare its predictions 
with those of the viscous ADAF models. 
As a most suitable candidate for such a comparison, Sgr A$^*$ is taken 
up here because it has been observed in many wave lengths as 
the nearest galactic nucleus and its spectrum has been reproduced 
many times by the successively advancing viscous ADAF models. 

In \S \ref{scaled}, we introduce the set of analytic solutions 
for resistive ADAFs in a suitably scaled form and discuss their 
basic characteristics.
The relevant radiation mechanisms and the methods of calculation 
of the fluxes are described in \S \ref{cal}. 
These schemes are applied to Sgr A$^*$ in \S \ref{result} and the 
results are discussed in comparison with those of the viscous ADAF 
models.  
Finally in \S \ref{sum}, we summerize the main results and discuss 
some related issues.

\section{RESISTIVE ADAF SOLUTION}\label{scaled}

As a basis of our calculation of spectra, we introduce here 
the set of analytic solutions constructing the resistive 
ADAF model. This set may be considered as a counterpart of 
that found by Narayan $\&$ Yi (1994, 1995a) in the viscous ADAF 
scheme, but it should be emphasized that the former is not a 
self-similar solution as the latter. 
In the resistive ADAF model, there are three basic quantities 
and one parameter: mass of the central black hole $M$, mass accretion 
rate $\dot{M}$, strength of the external magnetic field $\vert B_0\vert$ 
and half-opening angle of the flow $\Delta$, respectively. 
We introduce the following normalizations for these quantities and 
for the radial distance $R$: 
\begin{equation}
    m\equiv \frac{M}{10^{6}M_{\odot}}, \hspace{1em}
     \dot m\equiv \frac{\dot{M}}{\dot{M}_{\rm E}}, \hspace{1em}
       b_{0}\equiv \frac{\vert B_{0}\vert}{1G}, \hspace{1em}
        \delta \equiv \frac{\triangle}{0.1}, \hspace{1em}
	 r \equiv \frac{R}{R_{\rm out}},
\end{equation}
where $R_{\rm out}$ denotes the radius of the disk's outer edge. 
The Eddington accretion rate is defined by
  ${\dot M}_{\rm E} \equiv L_{\rm E}/( 0.1 c^{2})$, 
which includes the efficiency factor of $0.1$. 
Note that this definition of $\dot{M}_{\rm E}$  and the normalization 
factor for black hole mass are different from those in K00.
The latter is chosen here as 10$^6$ $M_{\odot}$ for the convenience 
of the discussion of Sgr A$^*$. 

In spherical polar coordinates, the radius of outer edge and 
the radial-part functions of relevant physical quantities (for 
their angular parts, see also K99, K00) are written as 
\begin{equation}
  R_{\rm out}=1.5 \times 10^{16}\ b_{0}^{-4/5}\dot{m}^{2/5}m^{3/5}
   \quad{\rm cm},
\end{equation}
\begin{equation}
   \vert b_{\varphi}(r)\vert=10\ \delta^{-1} b_{0} r^{-1} \quad{\rm G},
  \label{eqn:B}
\end{equation}
\begin{equation}
   v_{\rm K}(r)=0.9\times 10^{8}\ b_{0}^{2/5}\dot{m}^{-1/5}m^{1/5}r^{-1/2}
     \quad{\rm cm}\ {\rm s}^{-1},
\end{equation}
\begin{equation}
   P(r)=4.0\ \delta^{-2}b_{0}^{2}r^{-2}
     \quad{\rm dyne\ cm}^{-2},
\end{equation}
\begin{equation}
   T(r)=1.8\times 10^{7}\ b^{4/5}_{0}\dot{m}^{-2/5}m^{2/5}r^{-1}
     \quad{\rm K},
  \label{eqn:T}
\end{equation}
\begin{equation}
   \rho (r)=1.3\times 10^{-15}\ \delta^{-2}b_{0}^{6/5}
    \dot{m}^{2/5}m^{-2/5}r^{-1} \quad{\rm g}\ {\rm cm^{-3}},
\end{equation}
where $b_{\varphi}$ is the toroidal magnetic field,
$ v_{\rm K}$ is the Kepler velocity,
$P$ is the gas pressure, $T$ is the temperature (common to 
electrons and ions) and $\rho$ is the density. 
Owing to the non-negligible pressure term in the radial force 
balance, the toroidal velocity in the disk is reduced by a factor 
of $1/\sqrt{3}$ from the Kepler value. The surface density and 
the optical depth are given, respectively, by  
\begin{equation}
   \Sigma(r)=\Sigma = 4.1\ \delta^{-1}b^{2/5}_{0}\dot{m}^{4/5}
   m^{1/5} \quad{\rm g}\ {\rm cm^{-2}},
\end{equation}
\begin{equation}
\tau_{\rm es} (r) \simeq \frac{1}{2}\kappa_{\rm es}\Sigma	
     = 8.2\times10^{-1}\ \delta^{-1} b_{0}^{2/5}\dot{m}^{4/5}m^{1/5},
\end{equation}
where $\kappa_{\rm es}$ is the opacity for electron scattering.
Note that these are independent of $r$ in the present model. 

Fig.\ 1 shows a schematic picture of an accretion disk in a global
magnetic field, whose precise structure is described by the solution 
given above.
Otherwise uniform external magnetic field is twisted by the rotational 
motion of accreting plasma, and there develops a large toroidal 
magnetic field in the middle latitude region. 
The behavior of this component, especially within the geometrically 
thin accretion flow, is given as $-b_{\varphi}\tanh\xi$ in the 
resistive ADAF solution, where $\xi=(\theta-\pi/2)/\Delta$ is the 
normalized angular variable. 
Owing to the appearance of a global $b_{\varphi}$, angular momentum of 
the accreting plasma becomes able to be carried away by the magnetic 
stress to distant regions along the poloidal magnetic field. 
The extraction of angular momentum guarantees the inward motion of 
the plasma, which gradually becomes large until it reaches near 
the rotational velocity at around the inner edge of the accretion disk. 
Although the magnetic lines of force are also bent inwardly toward 
the center of gravitational attraction, it has been shown that the 
dominant component is the toroidal one. 
This component also plays an essential role in the plasma confinement 
toward the equatorial plane and keeps it geometrically thin through 
its magnetic pressure. 

Before going into the detailed discussion of the radiative 
processes, we briefly mention some similarities and differences 
in the basic features of the viscous and resistive ADAF models.
It is worth noting that, in spite of the essential difference 
in the mechanisms of angular momentum transport and energy dissipation, 
the predictions for quantities such as temperature, density and 
optical depth are quite similar in both models. 
The temperatures in both models are as high as a fraction of the 
virial temperature of ions. 
Indeed, this is the case in the viscous models, though the electron 
temperature may deviates from it in the inner regions (e.g., 
Narayan \& Yi 1995b), and so also in the resistive model as can be 
confirmed from the analytic expression of $T$ (K99, K00). 
Such a high temperature makes a sharp contrast with the case of 
standard $\alpha$-disks (e.g., Frank, King $\&$ Raine 1992). 
Further, for a sub-Eddington mass accretion rate, the optical depth is 
dominated by the electron scattering and is smaller than unity. 
These are therefore common features of sub-Eddington ADAF models 
as expected. 

As one of the main differences, it may be stressed that the magnetic 
field in the resistive ADAF model is an ordered magnetic field and is 
determined self-consistently in the model from a boundary value. 
Therefore, the strength of the field is not a parameter as in the 
viscous ADAF models.
The ordered magnetic filed extract angular momentum from the 
accreting plasma and confines it in a disk structure 
against the gas pressure. 
Gravitational energy is released in the disk as the Joule 
heating and also as compressional heating of the flow. 
In the above analytic model of resistive ADAFs, these energies 
are fully advected down the stream. 
We calculate the radiation from the disk as a small perturbation 
from this solution. 

Another distinction may be in the energy partition between 
the electron and ion components. 
The viscous ADAF models assume that the viscous dissipation, 
which is large at large radii, heats mainly ions. Since the efficiency 
of radiation cooling is very small for ions compared with electrons 
and since energy transfer to the electron component is estimated to be 
negligible (Manmoto et al. 1997) except in outer portions, the flow 
is fully advective in most portions. 
For the electron component, on the other hand, the radiative 
cooling is balanced by the advective heating, near the inner edge. 
The electron temperature, therefore, deviates downwards largely 
from the ion temperature thus realizing a two-temperature structure. 

In contrast to the viscous heating, the resistive dissipation 
becomes large at small radii and seems to preferentially heat 
the electron component as suggested by Bisnovatyi-Kogan \& Lovelace 
(1997). 
In this case, the temperature difference is expected to remain 
rather small because the heating is effective for effective 
radiator. 
In any case, the resistive ADAF model in its present version assume 
a common temperature to both components, for simplicity. 
The examination of its two-temperature version may belong to  
a future work.

\section{CALCULATION OF SPECTRUM}\label{cal}

As mentioned above, the radiation spectrum from a resistive ADAF 
is calculated based on the analytic solution introduced in the 
previous section. 
Back reactions of the radiation cooling to this fully advective 
solution are negligible as far as its fraction in the total cooling 
rate is small. 
This has been roughly checked in a previous paper (K00).  
The method of calculating radiation fluxes described in this 
section will be applied to Sgr A$^*$ in the next section. 
The observed spectrum of Sgr A$^*$ in the frequency range from 
radio up to X-ray range is successfully explained in the viscous 
ADAF models by the three processes, i.e., synchrotron radiation, 
bremsstrahlung and inverse Compton scattering (Narayan, Yi \& 
Mahadevan 1995; Manmoto et al.\ 1997; Narayan et al.\ 1998). 
Although there may be some other components such as the 
radio-frequency excess over the Rayleigh-Jeans spectrum and 
$\gamma$-ray peak both of which need separate explanations (see, e.g., 
Mahadevan 1999 for the former, and Mahadevan, Narayan \& Krolik 1997 
for the latter), we ignore these components here for simplicity. 

Among the above three processes, the Compton scattering is treated 
separately from the other processes of emission and absorption. 
Therefore, we divide the total flux into two parts: the flux 
due to bremsstrahlung and synchrotron process $F_{\nu}$ 
and that due to the inverse Compton process $F^{\prime}_{\nu}$. 
The obtained fluxes are both integrated over the entire 
surfaces (upper and lower ones) of a disk, and added up to obtain 
the luminosity per unit frequency $L_{\nu}$. 

Temperature in the flow is vertically isothermal in the present model. 
In calculating the emission and absorption processes, 
the flow is further assumed to be locally plane parallel. 
Solving the radiative diffusion equation at a given radius $R$, 
we obtain the flux of the unscattered photons $F_{\nu}$ 
emanating from one side of the disk (Rybicki \& Lightman 1979) as 
\begin{equation}
  F_{\nu} =\frac{2\pi}{{\sqrt 3}}B_{\nu}
             \left[1-\exp(-2{\sqrt 3}\tau_{\nu}^{*})\right],
  \label{eqn:Fnu}
\end{equation}
where $B_{\nu}$ is the Planck intensity and $\tau_{\nu}^{*}$ is 
the vertical optical depth for absorption, 
\begin{equation}
  \tau_{\nu}^{*}(R) \simeq \frac{{\sqrt \pi}}{2}\kappa_{\nu}R\Delta.
\end{equation}
Assuming the local thermodynamic equilibrium (LTE), we can express 
the absorption coefficient $\kappa_{\nu}$ at the equatorial plane 
in terms of the volume emissivities $\chi_{\nu}$'s for bremsstrahlung 
and synchrotron processes: 
\begin{equation}
 \kappa_{\nu}=\frac{\chi_{\nu,\rm br}+\chi_{\nu, \rm sy}}
   {4\pi B_{\nu}}.
\end{equation}
Thus, equation (\ref{eqn:Fnu}) includes not only the effect 
of free-free absorption but also of synchrotron 
self-absorption at low frequencies. 

As the distribution function for thermal electrons, we assume 
that of the relativistic Maxwellian (in its normalized form), 
\begin{equation}
 N_{e}(\gamma)d\gamma=\frac{\gamma^{2}\beta
  \exp(-\frac{\gamma}{\theta_{e}})}{\theta_{e} K_{2}(\frac{1}
  {\theta_{e}})}d\gamma,
 \qquad \theta_{e}=\frac{k_{\rm B} T_{e}}{m_{e}c^{2}},
\end{equation}
because ADAFs tend to have so high temperatures that electron 
thermal energy can exceed its rest mass energy. 
Here, $\gamma$ is the Lorentz factor, $k_{\rm B}$ is the 
Boltzmann  constant and $K_2$ is the 2nd modified Bessel function. 
Actually, we use this formula only in the calculation of Comptonized 
photon flux below, while in those of bremsstrahlung and 
synchrotron processes we follow the works of Narayan \& Yi (1995b), 
and Manmoto et al. (1997) where it is replaced by a numerical 
fitting function. 

\subsection{Bremsstrahlung}

At relativistic temperatures, we must take into account not only 
electron-proton but also electron-electron encounters. 
Therefore, the total bremsstrahlung cooling rate per unit volume 
is written as
\begin{equation}
  q^{-}_{{\rm br}}=q^{-}_{ei}+q^{-}_{ee},
\end{equation}
where the subscripts $ei$ and $ee$ denote the electron-ion and
electron-electron processes, respectively.
The explicit expressions of the cooling rates are as follows. 
For the electron-proton process, 
\begin{eqnarray}
 q^{-}_{ei}
   &=& 1.25\ n_{e}^{2}\sigma_{\rm T}c
     \alpha_{f}m_{e}c^{2}F_{ei}(\theta_{e})\nonumber \\
   &=& 1.48\times 10^{-22}n_{e}^{2}F_{ei}(\theta_{e})
   \quad {\rm ergs}\ {\rm cm}^{-3}\ {\rm s}^{-1},
\end{eqnarray}
where $n_e$ is the electron number density, $\alpha_{f}$ is the 
fine-structure constant and $\sigma_{\rm T}$ is the Thomson 
cross-section, and further 
\begin{eqnarray}
 F_{ei}(\theta_{e}) 
   &=& 4\left(\frac{2\theta_{e}}{\pi^{3}}\right)^{0.5}
    (1+1.781\theta_{e}^{1.34}) \qquad {\rm for} 
    \quad \theta_e <1, \nonumber \\
   &=& \frac{9\theta_{e}}{2\pi}\ [\ \ln(1.123\theta_{e}+0.48)+1.5] \nonumber \\  
    && \qquad {\rm for} \quad \theta_e>1.
\label{eqn:Bei}
\end{eqnarray}
For the electron-electron process, 
\begin{eqnarray}
 q^{-}_{ee}
 &=& n_{e}^{2}c r_{e}^{2}m_{e}c^{2}\alpha_{f}
  \ \frac{20}{9\pi^{0.5}}(44-3\pi^{2})\theta_{e}^{3/2}\nonumber \\
   &&\times (1+1.1\theta_{e}+\theta_{e}^{2}-1.25\theta_{e}^{5/2})
     \nonumber \\
 &=& 2.56 \times 10^{-22} n_{e}^{2}\theta_{e}^{3/2}
     (1+1.1\theta_{e}+\theta_{e}^{2}-1.25\theta_{e}^{5/2}) \nonumber \\
 &&   \quad {\rm ergs}\ {\rm cm}^{-3}\ {\rm s}^{-1} 
\end{eqnarray}
when $\theta_e < 1$, and 
\begin{eqnarray}
  q^{-}_{ee}
  &=& n_{e}^{2}c r_{e}^{2}m_{e}c^{2}\alpha_{f}24\theta_{e}(\ln 2
   \eta \theta_{e}+1.28)\nonumber \\
  &=& 3.40\times 10^{-22}n_{e}^{2}\theta_{e}(\ln 1.123
  \theta_{e}+1.28)\nonumber \\
  &&  \quad {\rm ergs}\ {\rm cm}^{-3}\ {\rm s}^{-1}
\end{eqnarray}
when $\theta_e > 1$. 
Here, $r_{e}=e^{2}/m_{e}c^{2}$ is the classical electron radius
and $\eta=\exp(-\gamma_{E})=0.5616$.

The emissivity per frequency is given by 
\begin{equation}
  \chi_{\nu,{\rm br}} = q^{-}_{{\rm br}}\bar{G}
    \ \exp\left(-\frac{h\nu}{k_{\rm B} T_{e}}\right)
    \quad {\rm ergs}\ {\rm cm}^{-3}\ {\rm s}^{-1}\ {\rm Hz}^{-1},
\label{eqn:Chi}
\end{equation}
where $h$ is the Planck constant and $\bar{G}$ is the Gaunt factor 
which is written (Rybicki $\&$ Lightman 1979) as 
\begin{eqnarray}
  \bar{G} 
    &=& \frac{h}{k_{\rm B}T_{e}}\left(\frac{3}{\pi}
        \frac{k_{\rm B} T_e}{h\nu} \right)^{1/2} 
    \qquad {\rm for} \quad \frac{h\nu}{k_{\rm B} T_e}>1, \nonumber \\
    &=& \frac{h}{k_{\rm B}T_{e}}\frac{\sqrt{3}}{\pi}
        \ \ln \left(\frac{4}{\zeta}\frac{k T_e}{h\nu} \right)
        \quad {\rm for}\quad \frac{h\nu}{k_{\rm B} T_e}<1.
\label{eqn:G}
\end{eqnarray}

The above cited formule contain a few minor defects. 
For example, the non-relativistic limit calculated for electron-ion 
process from equations (\ref{eqn:Bei}) and (\ref{eqn:G}) differs by 
about 35\% from the standard formula (Rybicki \& Lightman 1979). 
Equation (\ref{eqn:G}) assumes the same values of the Gaunt factor 
for both electron-electron and electron-ion processes. 
In spite of these defects, we adopt the above formule according to 
Narayan \& Yi (1995b) and Manmoto et al. (1997), considering that these 
are the best ones we can employ at present throughout the energy rage 
of our interest. 
The adoption of the same formule as in the previous calculations is 
also suitable for the purpose of comparison of the predictions of 
different models, such as the viscous and resistive ones.

\subsection{Synchrotron Emission}

Synchrotron emission is an essential process to produce 
the radio wave-length part of the spectra from optically thin 
ADAFs in AGNs. 
Especially in the resistive ADAF model, some information about 
the strength of the ambient magnetic field may be obtained 
from the process of spectral fitting. 
 
The optically-thin synchrotron emissivity by relativistic Maxwellian 
electrons is calculated from the formula (Narayan \& Yi 1995b; 
Mahadevan, Narayan \& Yi 1996),  
\begin{eqnarray}
 \chi_{\nu,{\rm sy}}
   &=& 4.43\times 10^{-30}\ \frac{4\pi n_{e}\nu}{K_{2}(1/\theta_{e})} 
\ I^{\prime}\left(\frac{4\pi m_{e} c \nu}{3 e B\theta_{e}^{2}}\right)\nonumber \\ 
& &\qquad {\rm ergs}\ {\rm cm}^{-3}\ {\rm s}^{-1}\ {\rm Hz}^{-1},
 \label{eqn:synChi}
\end{eqnarray}
where $e$ is the elementary charge and 
\begin{eqnarray}
  I^{\prime}(x)=\frac{4.0505}{x^{1/6}}
  \left( 1+\frac{0.4}{x^{1/4}}+\frac{0.5316}{x^{1/2}}\right)
  \exp(-1.8899x^{1/3}).
\end{eqnarray}
In equation(\ref{eqn:synChi}), the argument of $I^{\prime}$ is 
specified as 
\begin{eqnarray}
  x \equiv \frac{2\nu}{3\nu_{0}\theta_{e}^{2}},
  \quad \nu_{0} \equiv\frac{e\vert B\vert}{2\pi m_{e}c},
\end{eqnarray}
where $B$ is the local value of magnetic field for which we substitute 
$b_{\varphi}$. 

\subsection{Inverse Compton Scattering}

The soft photons whose flux is given by equation (\ref{eqn:Fnu}) 
are Compton scattered by the relativistic electrons in the flow. 
We adopt the rate equation of Coppi \& Blandford (1990) as the basis 
of our considerations. This equation applies to homogeneous, isotropic 
distributions. 
The first term on the right-hand side of their equation describes 
the rate of decrease in the photon's number density with a given 
energy owing to the scattering into other energies, while 
the second term does the increase owing to the scattering into 
this energy from other energies. 

In the situations of our interest, we can neglect the first term 
because the number density of Comptonized photons are small compared with 
that of the seed photons. 
Instead, we use the second term iteratively to calculate the effects 
of multiple scattering. 
The scattering occurs on the average when the condition 
$c\sigma_{\rm T}n_edt=1$ is satisfied, where $t$ is time and $n_e$ 
is the number density of electrons. 
The probability that such a condition is satisfied $j$-times 
before the photons come out of the surface may be given by 
the Poisson formula, 
\begin{eqnarray}
   p_j =\frac{\tau_{e}^j\ {\rm e}^{-\tau_{e}}}{j!}.
\end{eqnarray}
Then, the production rate for the photons with a normalized 
energy $\epsilon\equiv h\nu/m_e c^2$ is given by 
\begin{eqnarray}
 \lefteqn{ \frac{d n(\epsilon)}{c\sigma_{\rm T}n_e dt}
   = \sum_{j=1}^{\infty} p_j\ \int d\gamma_j\ldots d\gamma_1 
     \ N_e(\gamma_j)\ldots N_e(\gamma_1) } \nonumber \\
 & & \times \int d\epsilon_j \ldots d\epsilon_1 \nonumber \\ 
 & &     \ \Bigl[ P(\epsilon;\epsilon_j,\gamma_j) \ldots 
      P(\epsilon_2;\epsilon_1,\gamma_1)  
     \ R(\epsilon_j,\gamma_j) \ldots \nonumber \\
 & &  R(\epsilon_1,\gamma_1) 
     \ n_{\rm in}(\epsilon_1) \Bigl], 
 \label{eqn:Comp}
\end{eqnarray}
where $m_e$ is the electron mass and $n_{\rm in}$ is the number 
density of seed photons. 

The non-dimensional scattering rate $R(\epsilon,\gamma)$ including 
Klein-Nishina cross section $\sigma_{\rm KN}$ is written explicitly 
(Coppi $\&$ Blandford 1990) as 
\begin{eqnarray}
 R(\epsilon,\gamma)
   &=& \int_{-1}^1\frac{d\mu}{2}\ (1-\beta\mu)
       \ \frac{\sigma_{\rm KN}(\beta,\ \epsilon,\ \mu)}
       {\sigma_{\rm T}} \nonumber\\
   &=&\frac{3}{32 \gamma^{2}\beta \epsilon^{2}}
      \int^{2\gamma (1+\beta)\epsilon}_{2\gamma (1-\beta)\epsilon} dx
	\left[\left(1-\frac{4}{x}-\frac{8}{x^{2}}\right)
	 \ln(1+x)\right.\nonumber \\
    &&\left.+\frac{1}{2}+\frac{8}{x} -\frac{1}{2(1+x)^{2}}\right] 
      \qquad {\rm cm}^{3}\ {\rm s}^{-1}.
\end{eqnarray}
Scattered-photon distribution is denoted by 
$P(\epsilon;\epsilon^{\prime},\gamma)$ and, in the present calculation, 
approximated by a $\delta$-function (Lightman \& Zdziarski 1987, 
Fabian et al. 1986): 
\begin{eqnarray}
 P(\epsilon;\epsilon^{\prime},\gamma)=\delta\left(\epsilon
     -\frac{4\gamma^{2}}{3}\epsilon^{\prime}\right).
\end{eqnarray}
This is merely for simplicity and a more exact expression has been 
derived by Jones (1968) and corrected afterwards by Coppi \& 
Blandford (1990). 

Although the sum in equation (\ref{eqn:Comp}) runs to infinity, 
it seems appropriate to assume that photons which are scattered more
than certain times become saturated and obey the Wien distribution
$\propto \nu^3 \exp(-h\nu/kT_{e})$ (e.g., Manmoto et al. 1997). 
In view of the smallness of the optical depth in most sub-Eddington 
ADAFs ($ \tau_{\rm es} < 10^{-3}$, in the case of Sgr A$^*$), 
however, we truncate the power series in $\tau_{\rm es}$ at $j=2$ 
and ignore the saturation effect. 
After performing the integrations containing $\delta$-functions 
and transforming the photon number densities into fluxes by 
multiplying $ch\epsilon/2$ on both sides of equation (\ref{eqn:Comp}), 
we obtain 
\begin{equation}
  F_{\nu}^{\prime}(0) 
    = {\rm e}^{-\tau_{\rm es}}\ \left[\tau_{\rm es}F_{\nu}^{(1)}
      + \frac{\tau_{\rm es}^{\ 2}}{2}F_{\nu}^{(2)} \right], 
\end{equation} 
where once- and twice-scattered fluxes are given, respectively, by 
\begin{equation}
  F_{\nu}^{(1)} \equiv \int_1^{\infty}d\gamma_1
     \ N_e(\gamma_1)  
     \ R\left( \frac{3\epsilon}{4\gamma_1^{\ 2}}, \ \gamma_1 \right)
     \ F_{\rm in} \left( \frac{3\epsilon}{4\gamma_1^{\ 2}} \right),
\end{equation}
\begin{eqnarray}
  F_{\nu}^{(2)} 
   &\equiv& \int_1^{\infty}d\gamma_2\int_1^{\infty}d\gamma_1 
      \ N_e(\gamma_2)\ N_e(\gamma_1) \nonumber\\  
   & &\times\ R\left(\frac{3\epsilon}{4\gamma_2^{\ 2}},\ \gamma_2\right)
      \ R\left(\frac{3}{4\gamma_2^{\ 2}}\frac{3\epsilon}
      {4\gamma_1^{\ 2}},\ \gamma_1 \right) \nonumber \\ 
   & &\times\ F_{\rm in}\left(\frac{3}{4\gamma_2^{\ 2}}
      \frac{3\epsilon}{4\gamma_1^{\ 2}}\right)
\end{eqnarray}
with the definition $F_{\nu, {\rm in}}=(ch\epsilon/2)
n_{\rm in}(\epsilon) \equiv F_{\rm in}(\epsilon)$. 
As the incident flux $F_{\nu, {\rm in}}$ in the above expressions, 
the result from equation (\ref{eqn:Fnu}) should be used. 
The effects of low energy tail ($\beta<1$) in the electron 
distribution is neglected in performing the $\gamma$-integrals. 

\section{APPLICATION TO SAGITTARIUS A$^*$}\label{result}

In order to compare the resistive ADAF model with the current 
models of viscous ADAFs in their predictions of spectra from 
accretion flows, we apply the former model to Sgr A$^*$. 
The observed spectral data available so far have been compiled 
by Narayan et al. (1998). 
They assume that the interstellar column density is $N_{\rm H}=6 
\times 10^{22}\ {\rm cm}^{-2}$ and the distance to the Galactic 
center is $d=8.5\ {\rm kpc}$. 
In judging the accuracy of fittings between the calculated and 
observed spectra, a considerable weight has been put on the 
high resolution data points such as the VLBI radio (86 GHz) 
data and the $ROSAT$ X-ray data. 
It should be kept in mind, however, that the $ROSAT$ data may be 
interpreted as an upper limit because its resolution (PSPC) 
$\sim 20''$ is not considered as satisfactory and that other 
issues like the value of $N_{\rm H}$ are still under discussion.

We discuss the two cases in the resistive model, which will be 
called the compact-disk and the extended-disk models, respectively. 
These names come from the difference in extension of the disk 
which is represented by the radius ratio of the inner to the outer 
edges, $R_{\rm in}/R_{\rm out} = r_{\rm in}$. 
As confirmed below, this value is largely affected by the choice 
of position of the inner edge, $R_{\rm in}$.

  \subsection{Compact Disk Model}

According to the spirit of original resistive ADAF model, the inner 
edge in this case is determined by the magnetic flux conservation 
(K00). This gives an expression 
\begin{eqnarray}
R_{\rm in}=(1+\Delta^{-1})^{-2}R_{\rm out} 
          \simeq \Delta^{2} R_{\rm out},
\end{eqnarray}
where the last expression is valid only for thin disks ($\Delta \ll 1$). 
Note that this procedure is independent of the notion of the marginally 
stable orbit around black holes. 
The outer edge has been fixed, on the other hand, from the mass 
conservation as 
\begin{eqnarray}
 R_{\rm out}=\left(\frac{3GM{\dot M^{2}}}{B_{0}^{4}}\right)^{1/5}.
\end{eqnarray}

Fig.\ 2 shows the best fit spectrum in this model and the set of best 
fit parameters is 
\begin{eqnarray}
  & &M= 3.9 \times10^{5}\quad M_{\odot}, \nonumber \\ 
& &  {\dot M}=1.2\times10^{-4} {\dot M_{\rm E}}
   = 1.0\times10^{-6}M_{\odot}\ {\rm yr}^{-1},\qquad \nonumber \\
  & &\vert B_{0}\vert = 0.7 \quad {\rm G}, \qquad 
  \Delta= 0.14 \quad {\rm rad}. 
\end{eqnarray}
From these values, other quantities of our interest are fixed as follows: 
\begin{eqnarray}
  & &R_{\rm in} = 6.1 \times 10^{12} \quad {\rm cm}, \qquad 
  R_{\rm out} = 3.1 \times 10^{14} \quad {\rm cm}, \nonumber \\
  & &T=3.4 \times 10^{8}\ r^{-1} \quad{\rm K}, \qquad 
  \vert b_{\varphi}\vert = 5.0\ r^{-1} \quad{\rm G}, \nonumber \\
  & &\rho=1.8\times10^{-17}\ r^{-1} \quad {\rm g\ cm}^{-3},  \nonumber \\
  & & \tau_{es}=3.1 \times 10^{-4}.
\end{eqnarray}
Thus, it turns out that the inner edge of the present model is fairly 
large compared with the marginally stable orbit, $R_{\rm ms}=3.5 
\times 10^{11}$ cm, for a Schwarzschild hole of the above mass. 

The changes in the spectrum caused by varying central mass $M$, 
accretion rate $\dot{M}$, external magnetic field $\vert B_0\vert$ 
and disk's half-opening angle $\Delta$ are demonstrated in 
Figs. 3, 4, 5 and 6, respectively. 
The spectral features are anyway quite analogous to those predicted 
by the viscous ADAF models. 
The results of a detailed comparison between the resistive and viscous 
ADAF models will be discussed in the final subsection, based on the 
predicted spectral features. 

\subsection{Extended Disk Model}

In this model, the inner edge of the accretion disk is 
set at the radius of the marginally  stable circular orbit 
around a Schwarzschild black hole, 
\begin{eqnarray}
R_{\rm in}=R_{\rm ms}=3R_{\rm G}=\frac{6GM}{c^{2}},
\end{eqnarray}
where $R_{\rm G}$ is the gravitational radius of the hole. 
This choice is motivated by the expectation that at around 
this radius the infall velocity inevitably becomes of 
the order of the rotational velocity, (i.e., $\Re\sim 1$ where $\Re$ 
is the magnetic Reynolds number, see K99, K00). 
The definition of the outer edge is the same as in the compact disk 
model. 
Note that the above definition of the inner edge is adopted 
also in the viscous ADAF models. 

The best fit parameters in this model are 
\begin{eqnarray}
  & &M= 1.0\times10^{6} \quad M_{\odot}, \nonumber \\  
 & &   {\dot M}=1.3\times10^{-4} {\dot M_{\rm E}} 
     = 2.9\times10^{-6}\quad M_{\odot}\ {\rm yr}^{-1}, \nonumber \\
  & &\vert B_{0}\vert = 1.0 \times 10^{-6}\quad {\rm G}, \qquad 
     \Delta= 0.20 \quad {\rm rad}.
\end{eqnarray}
These are used to fix the values of various scaled quantities: 
\begin{eqnarray}
  & &R_{\rm in} = 8.9\times 10^{11} \quad {\rm cm}, \qquad 
  R_{\rm out} = 2.7\times 10^{19} \quad {\rm cm}, \nonumber \\
  & &T = 1.0\times 10^{4}\ r^{-1} \quad {\rm K}, \qquad 
  \vert b_{\varphi}\vert = 5.0\times 10^{-6}\ r^{-1} 
   \quad {\rm G}, \nonumber \\
  & &\rho = 5.9\times 10^{-25}\ r^{-1}
   \quad {\rm g}\ {\rm cm}^{-3},  \nonumber \\ 
 & & \tau_{es}= 1.3\times 10^{-6}.
\end{eqnarray}

The best fit curve is shown in Fig. 7. 
The changes in the spectrum caused by varying central mass $M$, 
accretion rate $\dot{M}$, external magnetic field $\vert B_0\vert$ 
and disk's half-opening angle $\Delta$ are demonstrated in 
Figs. 8, 9, 10 and 11, respectively. 
The spectral shapes are very different from those of the 
compact-disk case and of the viscous ADAF models. 
Synchrotron emission has a very wide peak and bremsstrahlung 
is negligibly small. 
The former fact is due to a high temperature at the inner edge 
(see sub-subsection 4.3.2) and the latter, to lower densities 
in the disk. 
The emission in the X-ray band is supported by the inverse Compton 
scattering from the radio band. 
The temperature near the outer edge falls even to such a small 
value that the assumption of complete ionization becomes invalid. 
Although the position of outer edge may seem to be irrelevant 
from a viewpoint of spectrum, it is nevertheless important also in 
this case as a fitting boundary of the inner magnetic field to the 
external one. 
The fitting predicts that the boundary value is comparable to 
the interstellar field (a few $\mu$G). 

The fitting both to 86 GHz and {\it ROSAT} X-ray data points is 
possible also in this model. However, it is clear that the fitting 
curve runs above the observed upper limits in the IR band. 
The fitting in the frequency range from 100 to 1000 GHz also becomes 
considerably poor compared with the case of compact disk. 
For these reasons, we judge that this model cannot reproduce 
the observed broadband spectrum of Sgr A$^*$. 
This fact suggests again that the inner edge of the accretion disk 
does not coincide with the marginally stable orbit. 
The wide range of the disk's radii which is obtained from this 
fitting implies that $\Re(R_{\rm out}) \sim 6\times 10^3$. 
Since $\Re(R)$ represents the ratio of toroidal to poloidal 
magnetic fields, most parts of the disk are very likely to be 
unstable to global MHD instabilities of helical type. 
For this reason too, we consider that the present case 
(i.e., $R_{\rm in} = R_{\rm ms}$) is quite unrealistic, at least, 
for Sgr A$^*$. 

\subsection{Viscous v.s. Resistive ADAFs}

  \subsubsection{Dependence on Black Hole Mass} 

The spectra calculated from ADAF models of both viscous and 
resistive types commonly have the saturated part at the lower 
ends of the spectra due to the synchrotron self-absorption. 
It is of great interest to see that the luminosity $\nu L_{\nu}$ 
of this part is essential to determine the mass of the central 
black hole, in both types 
of models. 
Especially, in the viscous model, the luminosity of this frequency 
part is determined almost only by the black hole mass. 
The reason is as follows. 

The temperature in ADAFs may be considered essentially as 
the ion virial temperature and hence decreases as $\sim R^{-1}$. 
Apart from a numerical factor due to a reduced Keplerian rotation, 
this is exactly true in the resistive model. 
This is also true in the viscous models for the main part of 
an accretion flow except in the inner region where the electron 
temperature deviates from the ion temperature and remains almost 
constant (e.g., Narayan \& Yi 1995b). 
Therefore, the contribution to the spectrum from each annulus of 
radius $R$ and width d$R$ is equal. 
Integrating these contributions up to the outer edge, we obtain 
$L^{\rm RJ}_{\nu} \propto T_{e}(R_{\rm in})R_{\rm in}R_{\rm out} 
=T_e(R_{\rm out})R_{\rm out}^2$, where $R_{\rm in}$ is 
the radius of the disk's inner edge in the resistive model 
and of the outer edge of the two-temperature region in the 
viscous models. 

We have $T_e R\propto m$ commonly to both types of ADAF 
models. 
Further, since radius scales as the gravitational radius 
in the case of viscous ADAFs, we obtain the mass dependence 
\begin{equation}
  L^{\rm RJ}_{\nu} \propto m^2 \qquad \mbox{(viscous ADAF)},
\end{equation} 
confirming the above statement. 
On the other hand, in the case of resistive ADAFs, we have 
\begin{eqnarray} 
  L^{\rm RJ}_{\nu} \propto b_0^{-4/5}\dot{m}^{2/5}m^{8/5} 
  \qquad \mbox{(resistive ADAF)},
\end{eqnarray}
where the dependences on the parameters other than $m$ have 
come from the expression of $R_{\rm out}$. 
In spite of these dependences, the mass dependence is essential 
also in this case. 
This is because the dependence on $\dot{m}$ is rather weak 
and the value of $b_0$ is strongly restricted from the position 
of the synchrotron peak (see the discussion below). 

  \subsubsection{Synchrotron Peak}

We estimate the synchrotron peak frequency following Mahadevan 
(1997), and examine its behavior in both viscous and resistive 
models.
For each annulus of radius $R$ and width d$R$, the synchrotron 
photons in the radio range up to a critical frequency $\nu_{\rm c}$ 
are strongly self-absorbed and result in the Rayleigh-Jeans spectrum. 
Therefore, the critical frequency of the spectrum is determined 
by equating the contributions to $L_{\nu}$ from optically thick 
and thin sides of the frequency: 
\begin{eqnarray}
 & &2\pi\frac{\nu_{\rm c}^{2}}{c^{2}}k_{\rm B}T_e(R) 2\pi R\ {\rm d}R \nonumber\\ 
  & &= 4.43\times 10^{-30}\frac{4\pi n_{e}\nu_{\rm c}}{K_{2}(1/\theta_{e})}
  I^{\prime}(x_{\rm c})\ 4\pi\Delta R^2\ {\rm d}R, 
\end{eqnarray}
where $x_{\rm c}$ is defined as 
$x_{\rm c}= 2\nu_{\rm c}/(3\nu_0\theta_e^2)$. 
Solving this equation, we can determine the value of $x_{\rm c}$ 
numerically (Appendix B of Mahadevan 1997). 
Provided that this value does not depend strongly on $R$, $\Delta$ and 
other parameters, we obtain 
\begin{eqnarray}
  \nu_{\rm c} = \frac{3}{2}\theta_{e}^{2}\nu_{0}x_{\rm c} 
         \propto T_e^{\ 2}(r)B(r). 
\end{eqnarray}
If the disk has uniform temperature and magnetic field, then 
the synchrotron peak is rather sharp and has a well-defined peak 
frequency at $\nu_{\rm c}$. 

When they vary with the radius $R$, however, substitution of 
the $r$-dependences of $T_{\rm e}$ and $B$ in both viscous and 
resistive ADAF models yield 
\begin{eqnarray}
  \nu_{\rm c} 
   &\propto&
   \alpha^{-1/2}(1-\beta)^{1/2}\dot{m}^{1/2}m^{-1/2}r^{-13/4} \nonumber\\
& &      \qquad \mbox{(viscous ADAF)}, \nonumber\\
   &\propto& \delta^{-1}b_0^{13/5}\dot{m}^{-4/5}m^{4/5}r^{-3}\nonumber\\
 & &      \qquad \mbox{(resistive ADAF)}.  
\label{eqn:nuc}
\end{eqnarray} 
This means that $\nu_{\rm c}$ is larger for smaller radii and the 
higher most cutoff is due to the inner edge. 
The position of peak of the superposed emission is then given as 
$\nu_{\rm p}=\nu_{\rm c}(r_{\rm p})$, where $r_{\rm p}$ is the 
radius whose contribution to the synchrotron emission is  most 
dominant. 
The fairly narrow peak obtained in the compact-disk case indicates 
that $r_{\rm p}$ is located near the inner edge and the global peak 
shape is determined mainly by the inner most region of the disk. 

On the contrary, the synchrotron peak becomes very broad and dull 
in the extended-disk case. 
We have confirmed that the low-frequency side of the broad peak 
is due to a superposition of the contributions from annuli of 
$R_{\rm ms} \sim 10R_{\rm ms}$. 
However, the dull shape on the high-frequency side of the peak may be 
mainly due to a resulting high temperature ($T\sim 3\times10^{10}$ K) 
at the smaller inner edge. 
Actually, owing to this high temperature and low densities near the 
inner edge, the synchrotron self-absorption becomes less important 
in the high-frequency radio band and the intrinsic shape of the 
synchrotron emission at the mildly relativistic temperature 
(Mahadevan et al. 1996) can appear on the high-frequency side. 

In any case, since $r_{\rm p}$ is a numerical factor, we can speak 
of the parameter dependences of the peak frequency $\nu_{\rm p}$ 
based on equation (\ref{eqn:nuc}). 
Note that the dependences on $m$ and $\dot{m}$ have different 
senses in the different ADAFs. 
The most important difference between the two models is that 
the dependence of $\nu_{\rm p}$ on the magnetic field is much more 
sensitive in the resistive model. 
Therefore, the field strength is determined more accurately there. 
All the predicted dependences on $m$, $\dot{m}$, $b_0$ and 
$\delta$ are qualitatively confirmed in Figs. 2 through 5. 
From the above considerations on the synchrotron peak, we think that 
the improvement of observational quality in submillimeter range is 
most important for obtaining more exact values of the disk parameters. 

  \subsubsection{Bremsstrahlung}

We shall try here to grasp the qualitative behavior of the contribution 
from bremsstrahlung according to the usual non-relativistic scheme. 
The contribution to a given frequency $\nu$ from optically thin plasma 
in an annular volume of width d$R$ is proportional to $\rho^2 T^{-1/2}
\exp[-h\nu/k_{\rm B}T]\ R^2{\rm d}R$. 
Apart from the exponential factor, we have 
\begin{eqnarray}
 \rho^2 T^{-1/2}R^2{\rm d}R 
   &\propto& \alpha^{-2}\dot{m}^2m r^{-1/2}{\rm d}r
        \qquad \mbox{(viscous ADAF)}, \nonumber \\
   &\propto& \delta^{-4}b_0^{-2/5}\dot{m}^{11/5}m^{4/5} r^{1/2}{\rm
   d}r \nonumber\\
     & &   \qquad \mbox{(resistive ADAF)}. 
\end{eqnarray}
Therefore, the relative importance of the inner and outer parts of 
a disk can be seen from the ratio, 
\begin{equation}
  f \equiv 
      \frac{\rho^2_{\rm in}T^{-1/2}_{\rm in}R^2_{\rm in}
      \exp[-h\nu/k_{\rm B}T_{\rm in}]}
      {\rho^2_{\rm out}T^{-1/2}_{\rm out}R^2_{\rm out}
      \exp[-h\nu/k_{\rm B}T_{\rm out}]} 
    \simeq \zeta^{\pm 1/2}\exp[h\nu/k_{\rm B}T_{\rm out}],
\end{equation}
where $\zeta\equiv R_{\rm out}/R_{\rm in} = r_{\rm in}^{-1}$, and 
the upper and lower signs in its exponent are for the viscous 
and resistive ADAFs, respectively. 

It is evident from the above ratio that, in the viscous ADAF, the 
contribution from the inner disk is always dominant (i.e., $f>1$) 
irrespective of the frequency $\nu$. 
In the resistive ADAF, however, it depends on the frequency, 
so that we introduce the critical frequency $\nu_{\rm c}^{\prime}$ 
by the relation $f=1$. 
This yields 
\begin{equation}
  \nu_{\rm c}^{\prime} = \frac{\ln\zeta}{2}
                         \ \frac{k_{\rm B}T_{\rm out}}{h}.
\end{equation}
Then, the inner part contributes to the frequency range 
$\nu > \nu_{\rm c}^{\prime}$ and the outer part, 
to $\nu < \nu_{\rm c}^{\prime}$. 
In fact, the critical frequency roughly coincides with the peak 
frequency of the bremsstrahlung. 
The luminosity above $\nu_{\rm c}^{\prime}$ can be roughly estimated, 
by putting ${\rm d}r\sim r \sim r_{\rm in}$, as 
\begin{eqnarray}
 L_{\nu}^{\rm br} 
   &\propto& \alpha^{-2}\dot{m}^2m 
        \qquad \mbox{(viscous ADAF)}, \nonumber \\
   &\propto& \delta^{-1}b_0^{-2/5}\dot{m}^{11/5}m^{4/5} 
        \qquad \mbox{(resistive ADAF)}, 
\end{eqnarray}
because $r_{\rm in}$ is a numerical constant and, in particular, equal 
to $\delta^2$ in the resistive ADAFs. 

In the viscous ADAF models and in the compact-disk case of the 
resistive ADAF, the contributions from bremsstrahlung cause an X-ray 
bump in each predicted spectrum. 
The dependences of this peak on the parameters $m$ and $\dot{m}$ 
are qualitatively confirmed in Figs. 2 and 3, but those on $b_0$ 
and $\delta$ are somewhat different from the above prediction, 
indicating a limitation of such a crude estimate as the above. 
The critical frequencies calculated from the best fit values 
for the compact and extended disks are $1.4\times10^{19}$ Hz and 
$1.8\times10^{15}$ Hz, respectively. 
The former value is in good agreement with the peak of the reproduced 
spectrum. 
In the extended-disk case, the contribution from the bremsstrahlung 
is negligibly small because the density throughout the disk becomes 
too small, and the X-ray range of the spectrum is explained by 
the once-scattered Compton photons.

\section{Summary $\&$ Discussion}\label{sum}

To summarize the examinations in the previous section, 
both viscous and resistive ADAF models can explain 
the observed spectrum of Sgr A$^*$ equally well. 
In spite of large differences in the basic mechanisms working 
in both models, the calculated spectra are quite similar, 
except for the extended-disk case in the resistive model. 
This fact suggests that also the resistive ADAF model is 
quite powerful in explaining the behavior of other low luminosity 
AGNs (Narayan, Mahadevan \& Quataert 1998).
In addition to these analogous aspects, the resistive model 
seems to have a possibility to explain such an essentially 
different situation as appeared in the extended disk case. 
In any case, when the presence of an ordered magnetic 
field should be taken seriously in some AGNs or in some 
stellar-size black holes then the resistive ADAF model, 
whose predictions on the radiation spectra are examined in 
this paper, will serve the purpose. 

One of the most remarkable features of the ADAF models 
is that the mass of the central black hole seems to be 
determined only from the fitting to the self-absorbed part 
of the observed spectrum. 
In the case of Sgr A$^*$, the resistive ADAF model (hereafter 
restricting to the case of compact disk) predicts the central 
mass of $3.9\times 10^5\ M_{\odot}$ while the viscous 
models predict $1.0\times10^6\ M_{\odot}$ (Manmoto et.al 1997) 
and $2.5\times 10^6\ M_{\odot}$ (Narayan et al. 1998). 
The accuracy of the fittings for other disk parameters 
than the black hole mass will be greatly improved by the 
precise determination of the position and height of the 
synchrotron peak from observations. 

The black hole mass predicted by the resistive ADAF model is 
evidently smaller compared with the predictions of the viscous 
ADAF models. 
The latter values are consistent with the dynamically reduced 
value of $2.5\times10^6\ M_{\odot}$ (Haller et al.\ 1996; 
Eckart \& Genzel 1997), which may be considered as an upper limit 
for the black hole mass. 
In the history of viscous ADAF models, the predicted black-hole 
mass was as small as $7\times10^5\ M_{\odot}$ (Narayan et al.\ 
1995). Afterwards, by the inclusion of compressive heating, 
it becomes consistent with the dynamical mass. 
Since this change is mainly due to the decrease in electron 
temperature (Narayan et al.\ 1998), the prediction of the resistive 
ADAF model may also be increased if the development of its 
two-temperature versions results in a lower electron temperature. 
As for the compressive heating, it is already included in the 
resistive model.  

In spite of the resemblance in the predicted spectral shape, 
there are of course many differences in the predictions of the 
viscous and resistive models. 
The precise dependences on the relevant quantities of the 
luminosities of the self-absorbed part, the synchrotron peak and 
the X-ray bump are different. 
Especially, the dependence of the synchrotron peak-frequency 
on the strength of magnetic field is much stronger for 
the resistive model.  
The essential difference in the geometry of an accretion flow may 
be in the radius of the inner edge rather than in its vertical 
thickness. 
The prediction of the resistive ADAF model for the inner-edge 
radius of the disk around Sgr A$^*$ is $\sim 20R_{\rm ms}$, 
instead of the radius of marginally stable circular orbit 
$R_{\rm ms}$. 
Although this result justifies the neglection of the general 
relativistic effects in our treatment, various questions may 
be raised about the behavior of infalling plasmas. 
As for this point we only present an idea below from a 
viewpoint of global consistency, because its detailed analyses 
are beyond the scope of this paper. 

Fig.\ 1 shows an overview of the flow and magnetic field 
configurations (see K00, for more details). 
The accretion flow would be decelerated near the inner edge 
by the presence of a strong poloidal magnetic field which is 
maintained by the sweeping effect of the flow. 
As a result, a certain fraction of the accreting plasma will 
be turned its direction to go along the poloidal field lines, 
although the remaining fraction may fall into the central black 
hole. 
If the poloidal current driven in the accretion disk can 
close its circuit successfully around distant regions and along 
the polar axis, a set of bipolar jets will be formed (Kaburaki 
$\&$ Itoh 1987). 
Even if the mechanism for formation of jet does not work well, 
the plasma within the inner edge is likely to extend to the polar 
regions. 

The presence of the plasma within the inner edge of an accretion 
disk and near the polar axis can be a possible source of the excess 
above the self-absorbed slope in radio band of the observed spectrum. 
Very recent VLBI observations of Sgr A$^*$ (Krichbaum et al. 1998; 
Lo et al. 1998) report that its intrinsic sizes in the east-west 
direction at 215 GHz and 68GHz are about 20 $R_{\rm G}$ (with 
$M=2.5\times 10^6M_{\odot}$). 
A half of this size (i.e., its radius) is just comparable to 
the size of the inner edge $\sim 60R_{\rm G}$ of our model fitting 
with $M=4\times10^5M_{\odot}$. 
However, the value of the black hole mass estimated from the spectral 
fitting may be increased if there is a possibility for Sgr A$^*$ 
to have a wind-type mass loss from the surfaces of its disk (such 
possibilities have been noted for various types of objects by, e.g., 
Blandford \& Begelman 1999; Di Matteo et al. 1999; Quataert \& 
Narayan 1999). 
In such a case, the VLBI component becomes smaller than the size of 
the inner edge.

From the standpoint of the resistive ADAF, therefore, the above 
observations should be interpreted as suggesting the presence of 
a compact structure which is comparable to or smaller than the 
inner-edge radius of the accretion disk. 
In view of the vertical elongation of this component reported by 
Lo et al. (1998), this structure is very likely to be the root of 
a jet as suggested by them. 
This picture is very consistent with the view described above in 
relation to Fig.\ 1. 
In this case, however, the location of the self-absorbed slope in 
the $\nu$-$\nu L_{\nu}$ diagram should be slightly shifted towards 
the higher-frequency side so that the VLBI data points can be 
regarded as an excess from the disk's contribution. 
\\

One of the authors (M.\ K.) would like to thank Tadahiro Manmoto 
for many valuable comments on the viscous ADAF models. 
He is also grateful to Umin Lee for his suggestions on some numerical 
technics.

\newpage  

\begin{figure}
  \plotone{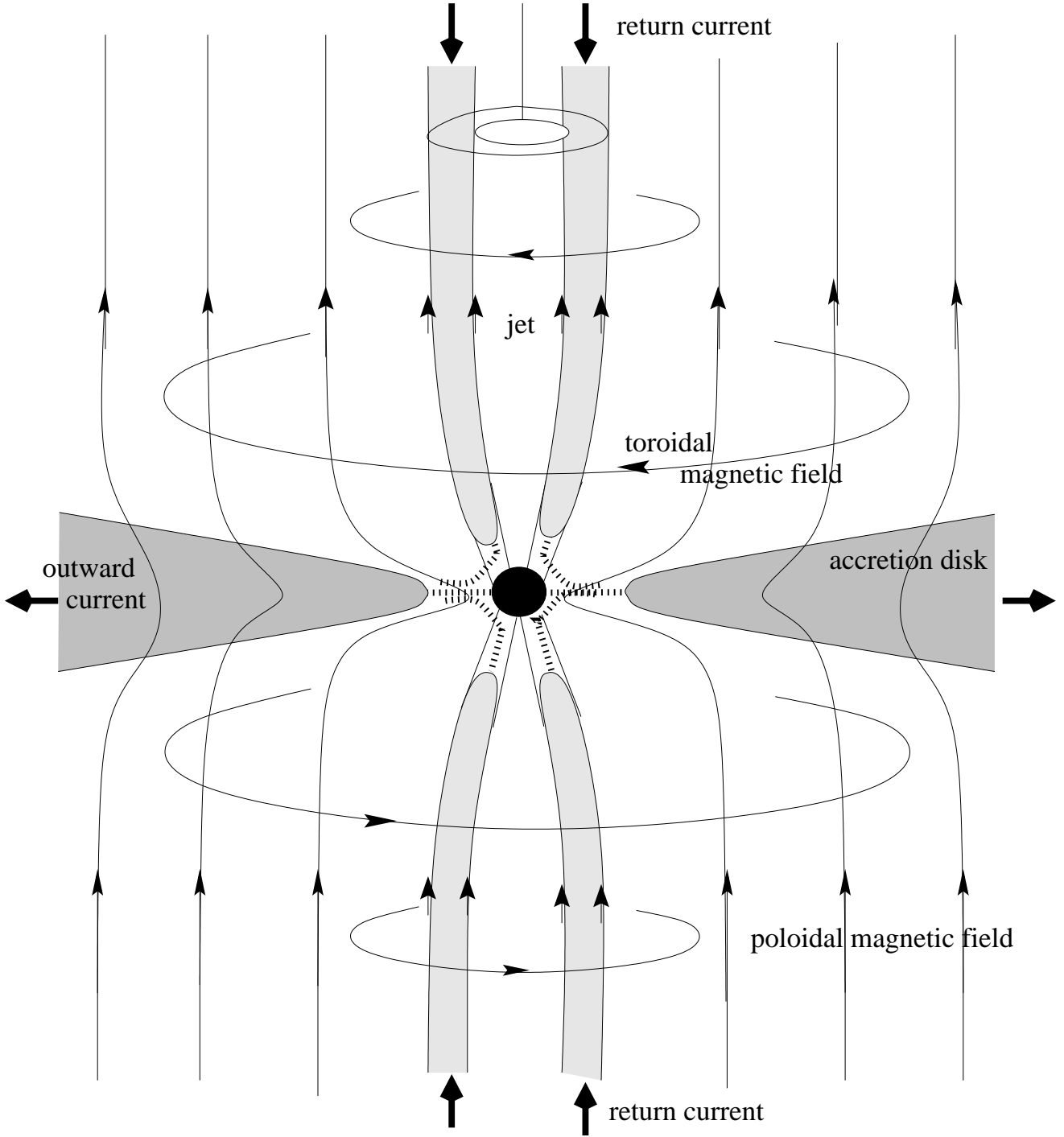}
\caption
{Schematic drawing of the global geometries of magnetic 
field and plasma flow. 
A poloidally circulating current system (${\bf j}_{\rm p}$) driven 
by the rotational motion of accreting plasma generates a toroidal 
magnetic field $b_{\varphi}$ in addition to a nearly uniform 
external field. 
The presence of this toroidal field outside the disk guarantees 
the magnetic extraction of angular momentum from the disk. 
This field also acts to confine the accreting flow toward the 
equatorial plane and has a tendency to collimate and accelerate 
the plasma in the polar regions. 
If the condition is favorable, the plasma in the polar regions 
may form a set of bipolar jets. 
In this paper, however, we focus our attention on the radiation 
spectrum from the accretion disk only. }
\end{figure}

\begin{figure}
  \plotone{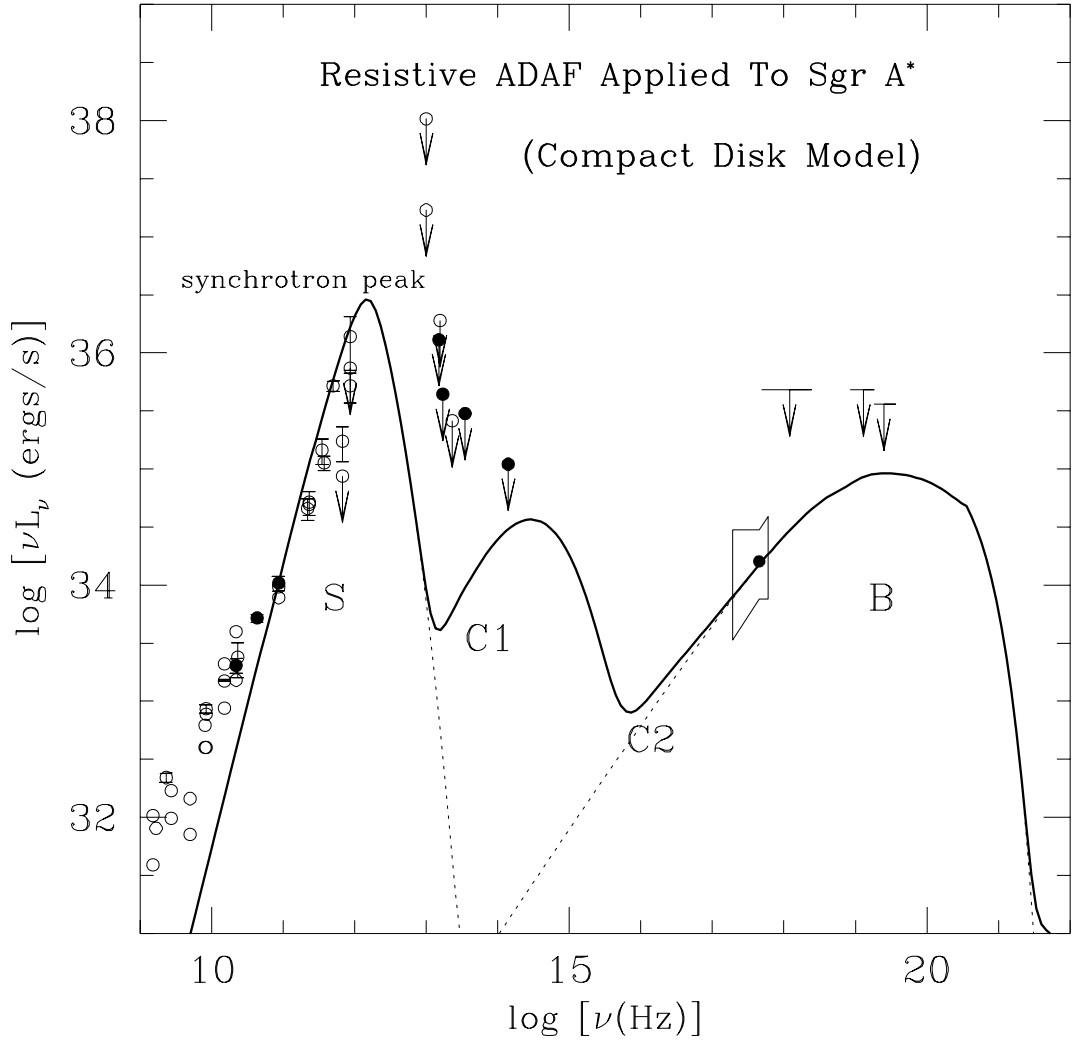}
  \caption
{The best fit spectrum of Sgr A$^{*}$ in the compact 
disk model. 
The resulting physical quantities are 
$M= 3.9\times10^{5}\ M_{\odot}$,
${\dot M}=1.2\times10^{-4}\ \dot{M}_{\rm E},$
$\vert B_0\vert = 0.7\ {\rm G}$ and $\Delta= 0.14\ {\rm rad}$. 
The four peaks indicated by S, C1, C2 and B denote 
those due to synchrotron emission, once- and twice-scattered 
(although it is almost buried) Compton photons and bremsstrahlung, 
respectively. 
The data points are the same as those compiled by Narayan 
et al. (1998).}
\end{figure}

\begin{figure}
 \plotone{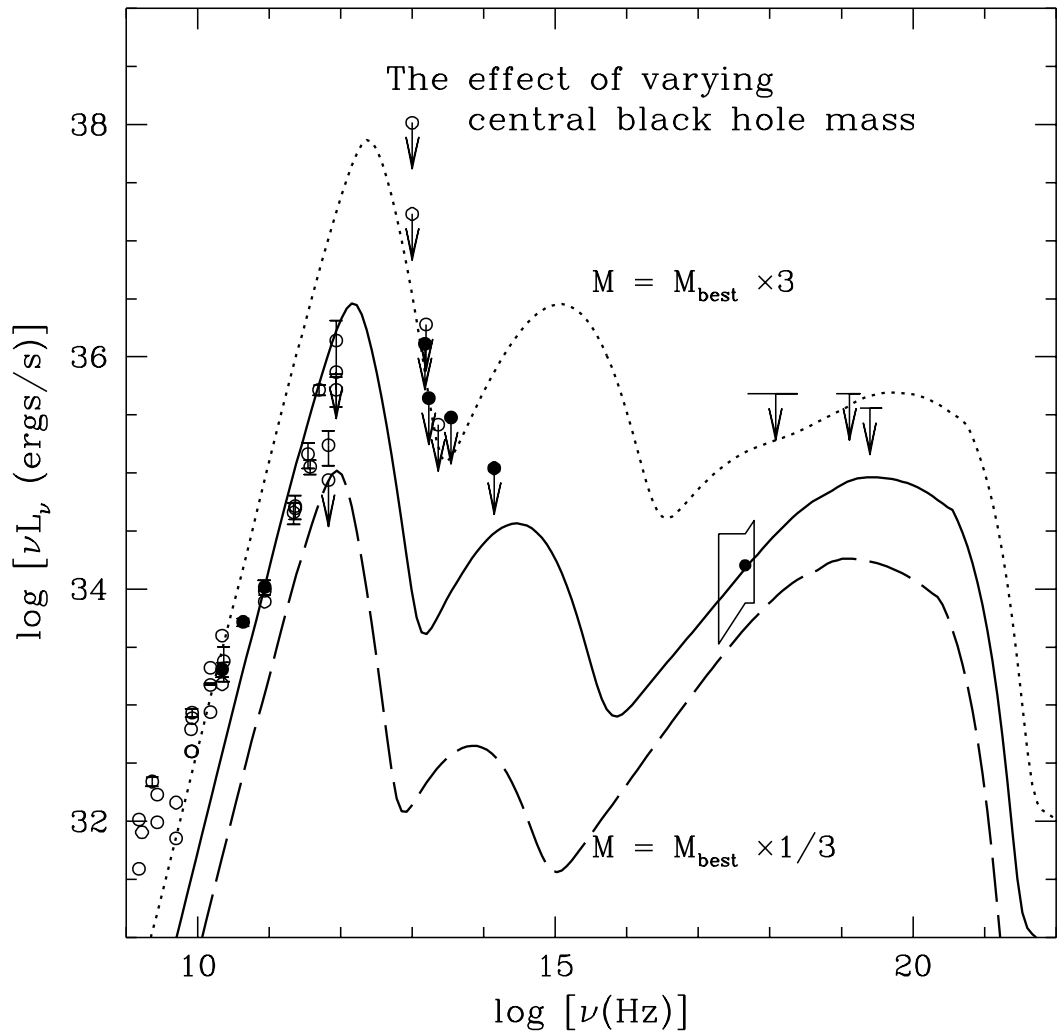}
\caption
 {With increasing central mass, the luminosity $\nu L_{\nu}$ 
increases globally, so that the self-absorbed part shifts upward 
and the synchrotron peak-frequency sifts to higher frequencies. 
The direction of the shift of the synchrotron peak is different 
from the results of viscous ADAFs (Manmoto et al. 1997, 
Narayan et al. 1998). }
 \end{figure}

  \begin{figure}
  \plotone{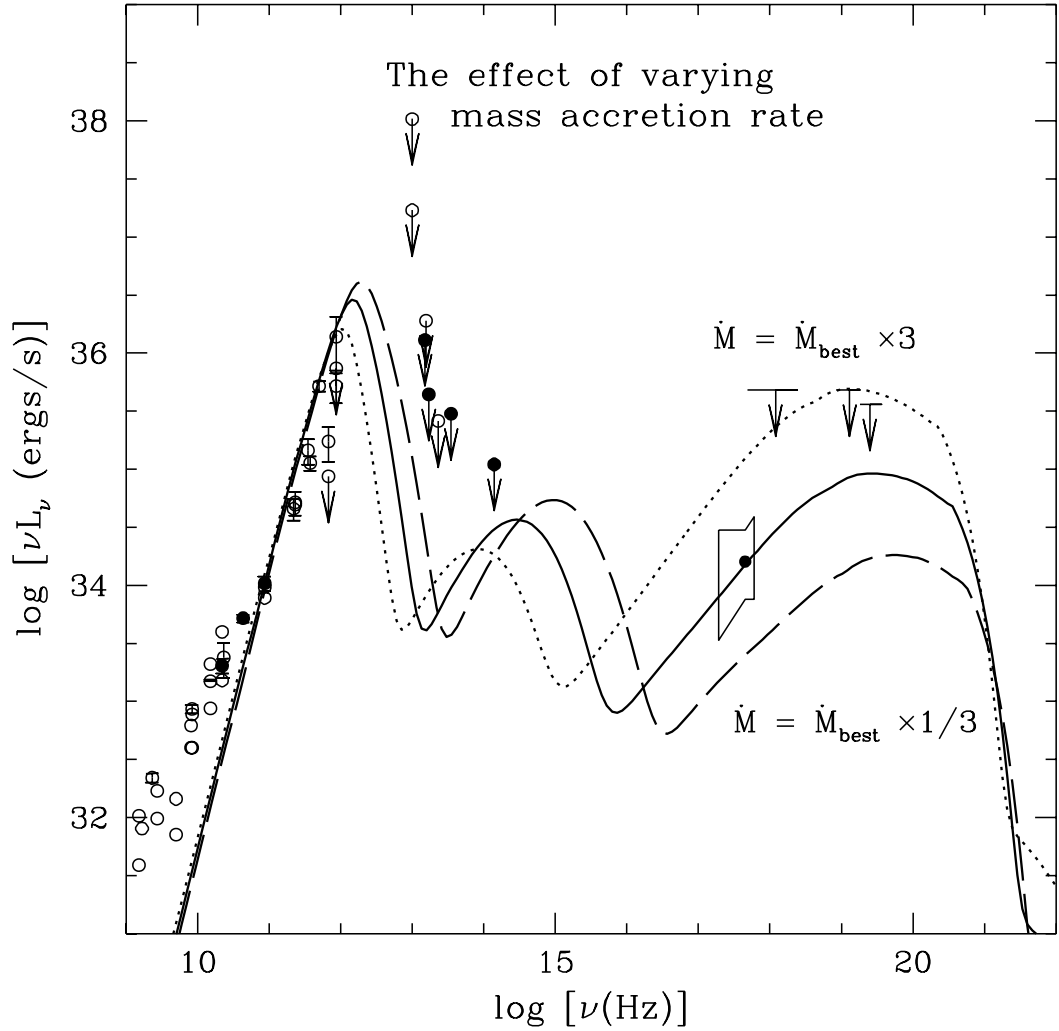}
  \caption
{$\dot{M}$-dependence of the spectrum. 
With increasing mass accretion rate, the synchrotron peak 
shifts to lower frequencies while the self-absorbed luminosity 
is hardly affected. 
The former behavior is opposite to that of viscous ADAFs. 
The X-ray bump increases with $\dot{M}$.}
  \end{figure}

  \begin{figure}
  \plotone{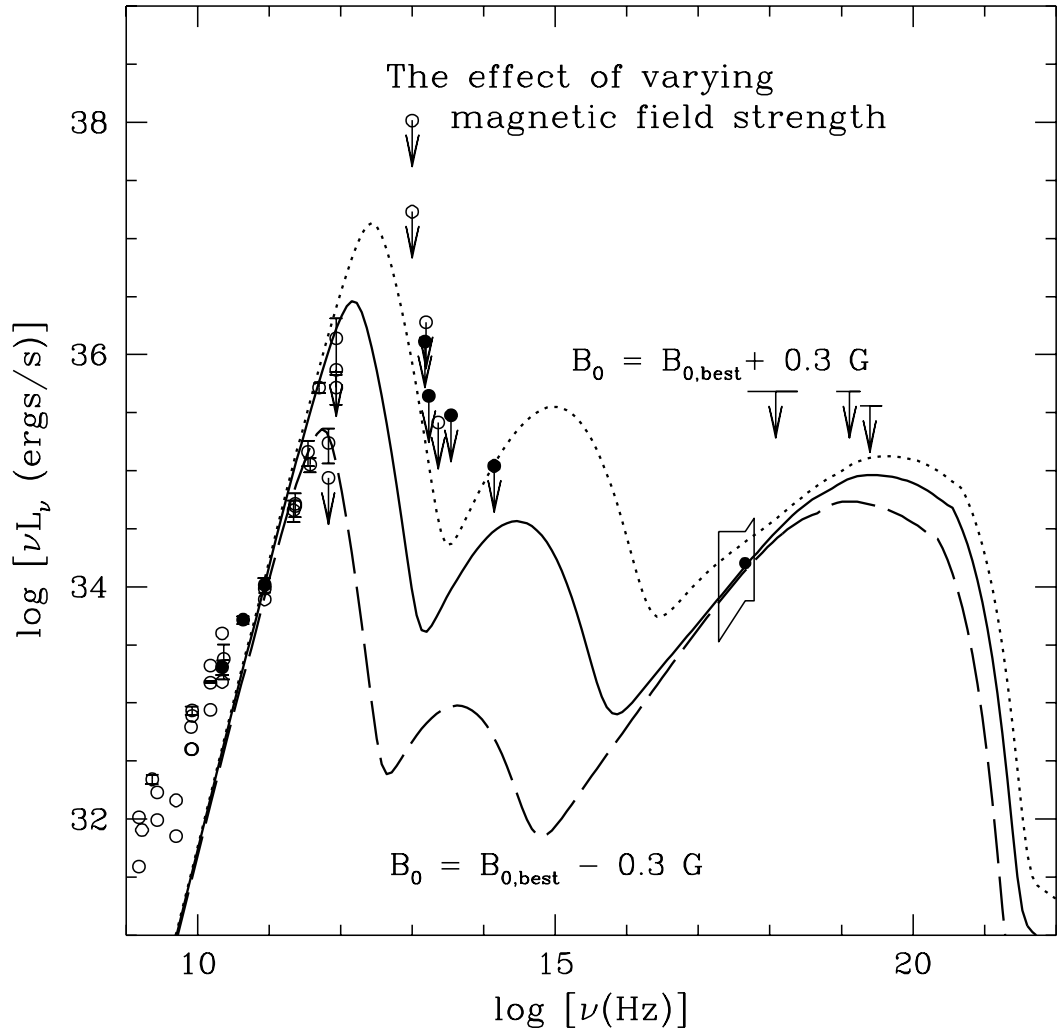}
 \caption
{$\vert B_0\vert$-dependence of the spectrum. 
With increasing magnetic field strength, the synchrotron peak shifts 
to higher frequencies while the self-absorbed luminosity is 
hardly affected. 
The X-ray bump increases with increasing $\vert B_0\vert$, 
suggesting a limitation of the crude estimate of $L_{\nu}^{\rm br}$ 
given in the text. }
  \end{figure}

\begin{figure}
  \plotone{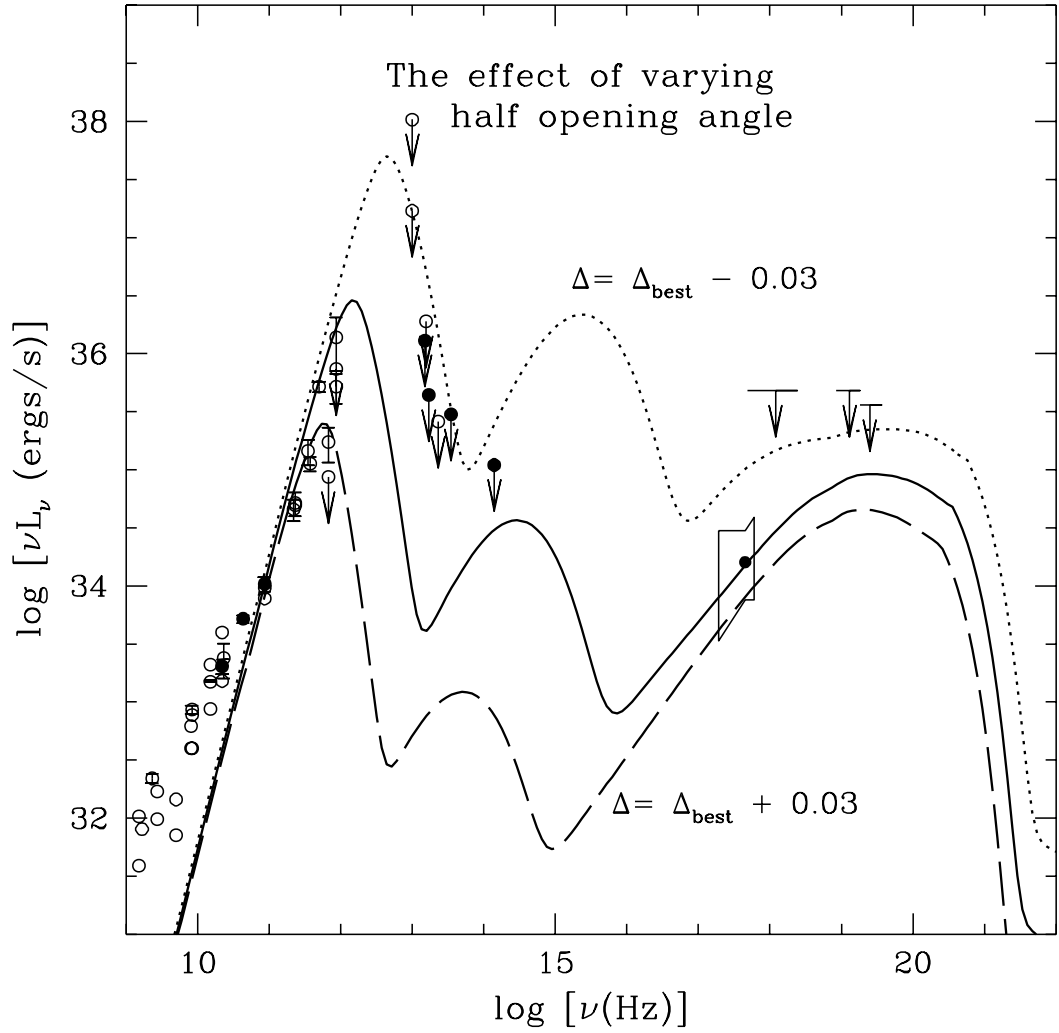}
 \caption
{$\Delta$-dependence of the spectrum. 
With increasing half-opening angle, the synchrotron peak 
shifts to lower frequencies while the self-absorbed luminosity 
is hardly affected. 
The dependence of the X-ray bump on $\Delta$ again suggests 
a limitation of the crude estimate given in the text. }
\end{figure}

\begin{figure}
  \plotone{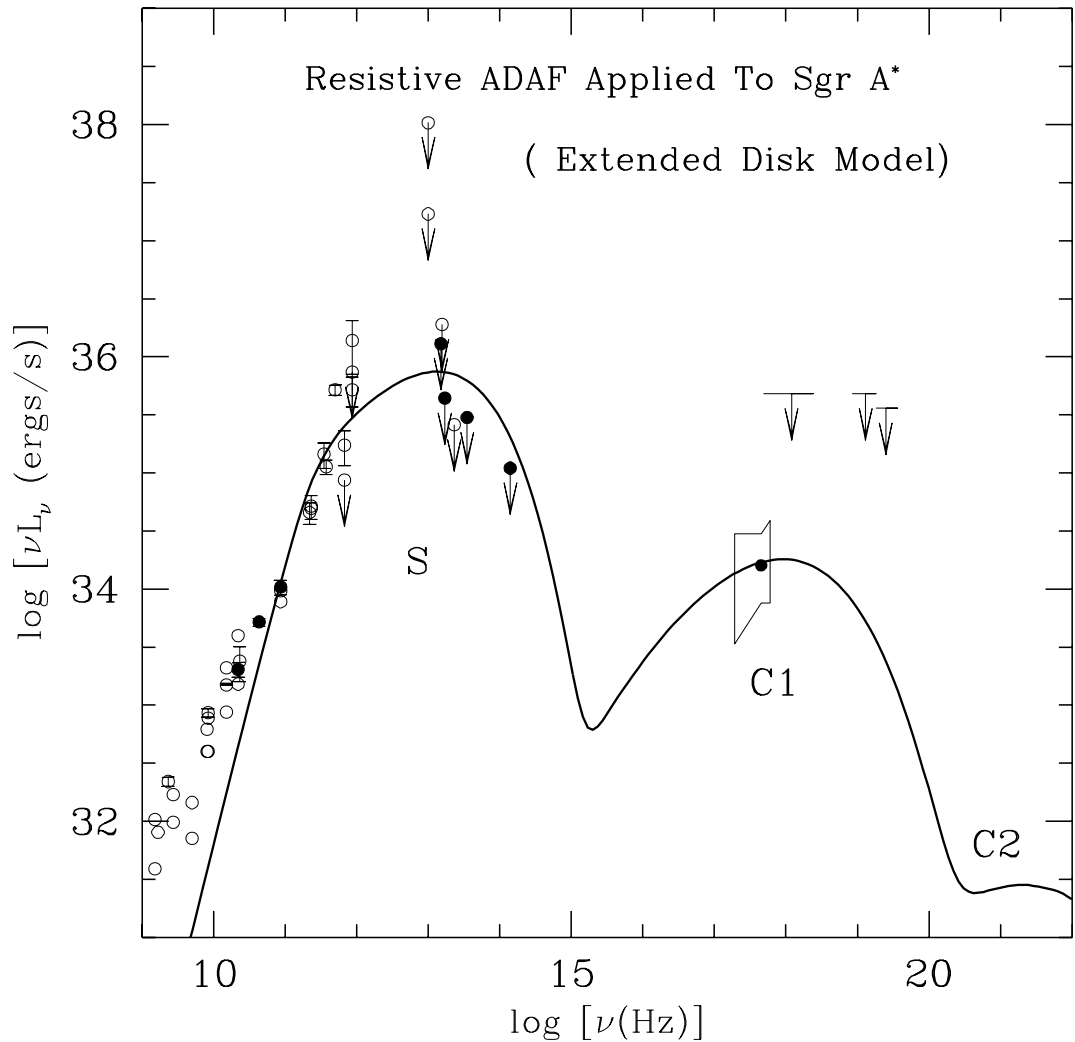}
 \caption
{The best fit spectrum of Sgr A$^{*}$ in the 
extended disk model. 
The resulting physical quantities are 
$M= 1.0\times10^{6}\ M_{\odot}$, 
${\dot M}=1.3\times10^{-4}\ {\dot M_{\rm E}}$, 
$\vert B_{0}\vert = 1.0 \times 10^{-6}\ {\rm G}$,
$\Delta= 0.20\ {\rm rad}$. 
Among the four peaks which are seen in the case of compact 
disk, that of bremsstrahlung has disappeared owing to 
the low densities in the extended disk, so that the X-ray bump 
is explained by once-scattered Compton photons. 
The fitting both to 86 GHz and {\it ROSAT} X-ray data points is 
possible also in this model. 
However, the fitting in the frequency range from 100 to 1000 GHz 
becomes considerably poor compared with the case of compact disk, 
and it cannot be reconciled with the observed upper limits 
in the IR band. 
For these reasons, we judge that this model cannot reproduce 
the observed broadband spectrum of Sgr A$^*$. 
}
\end{figure}

\begin{figure}
  \plotone{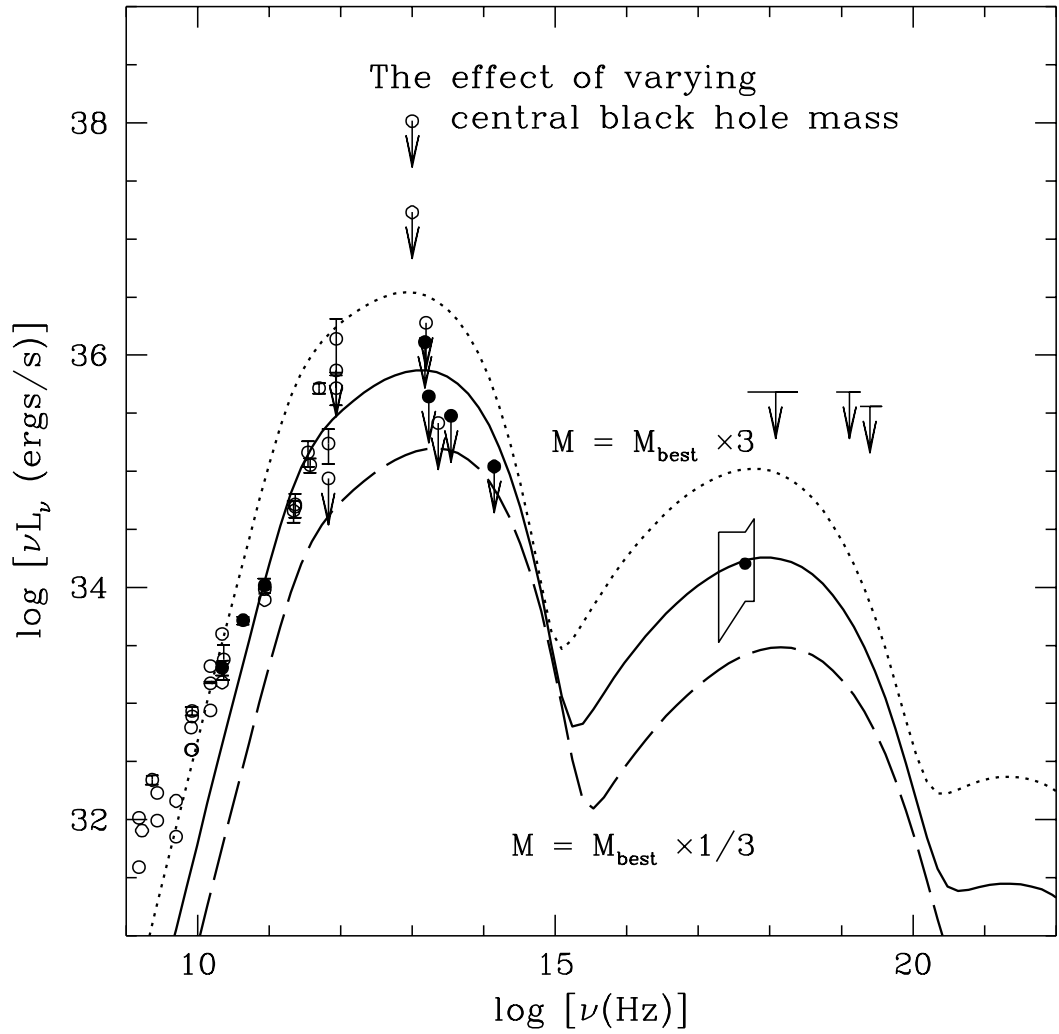}
\caption
{$M$-dependence of the spectrum. 
The tendency of the changes caused by varying $M$ is almost 
the same as in the case of compact disk. }
  \end{figure}

  \begin{figure}
  \plotone{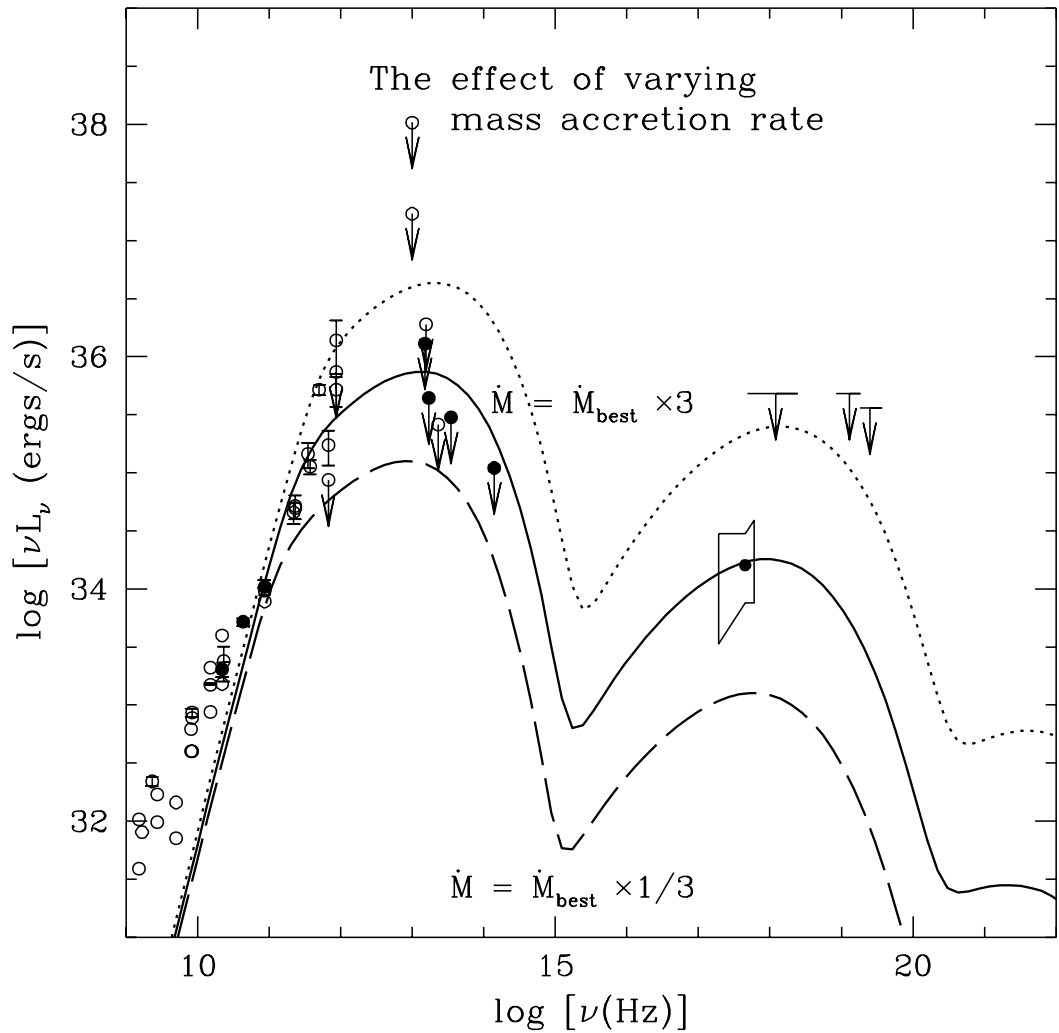}
 \caption
{$\dot{M}$-dependence of the spectrum. 
With increasing mass accretion rate, the synchrotron peak 
shifts to higher frequencies while the self-absorbed luminosity 
is hardly affected. 
The former tendency is different from the case of compact disk.}
   \end{figure}

  \begin{figure}
  \plotone{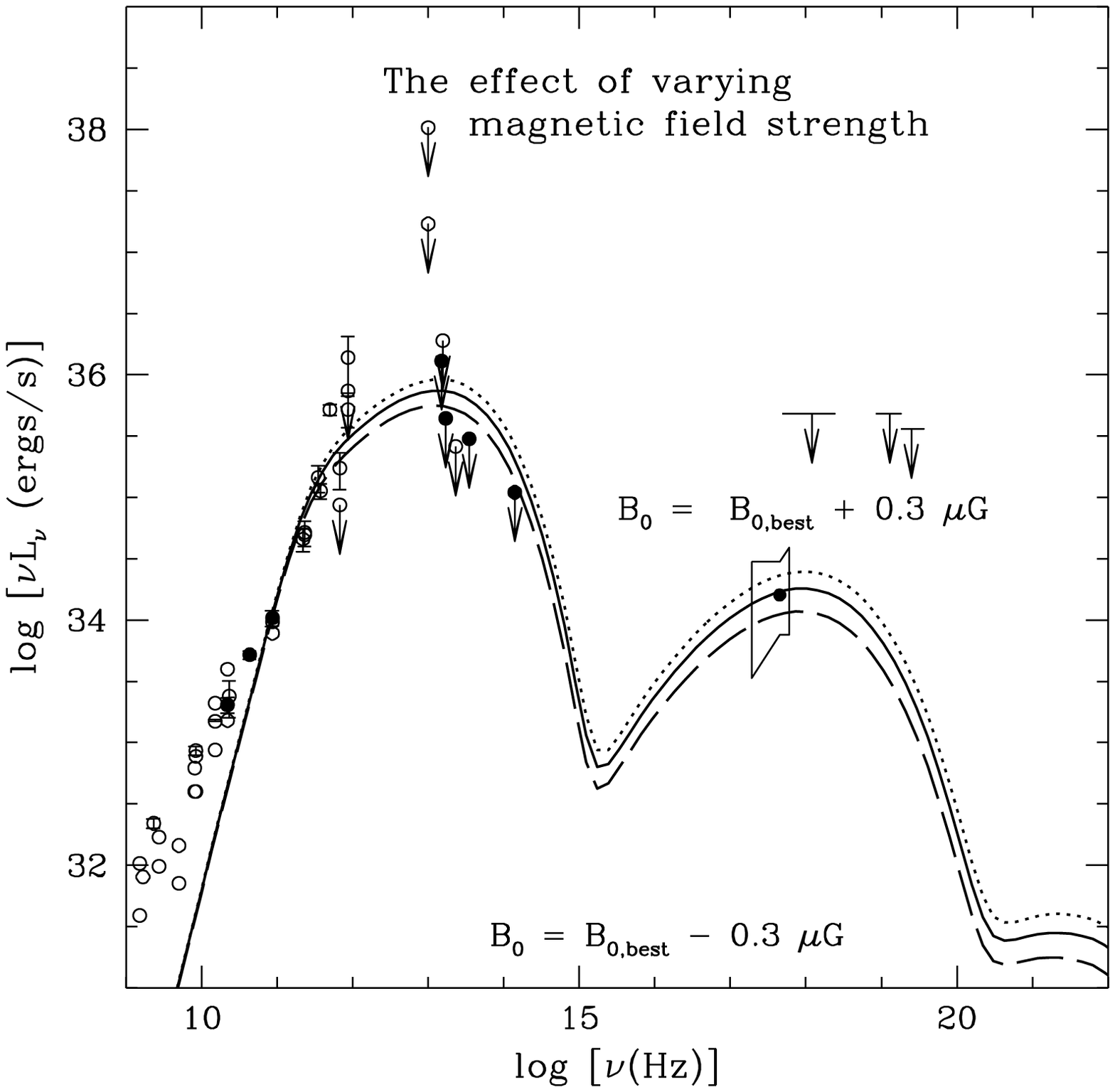}
\caption
{$\vert B_0\vert$-dependence of the spectrum. 
The tendency is almost the same as in the case of compact disk.}
  \end{figure}

\begin{figure}
  \plotone{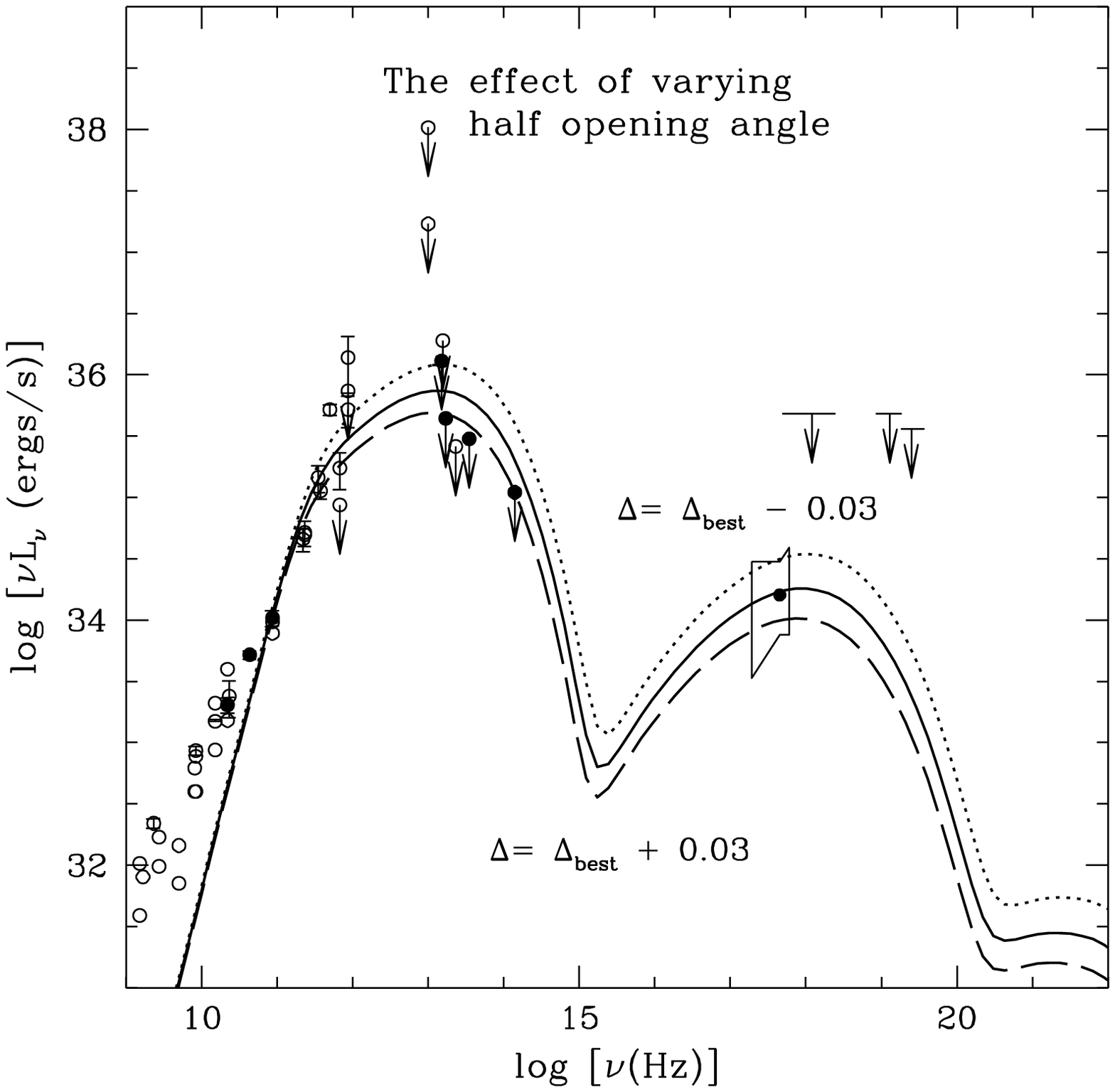}
\caption
{$\Delta$-dependence of the spectrum. 
The tendency is almost the same as in the case of compact disk.}
 \end{figure}


\begin{references}

 \reference{1}
{Abramowicz, M.\ A., Chen, X., Kato, S., Lasota, J.\ -P.,   
\& Regev, O.  1995, \apjl, 438, L37}
 \reference{1}
{Bisnovatyi-Kogan, G.\ S., \& Lovelace, R.\ V.\ E. 1997, 
\apjl, 486, L43 }
 \reference{1}
{Blandford, R.\ D., \& Begelman, M.\ C. 1999, \mnras, 303, L1}
 \reference{1}
{Coppi, P.\ S., \& Blandford, R.\ D. 1990, \mnras, 245, 453}
 \reference{1}
{Di Matteo, T., Fabian, A.\ C., Rees, M.\ J., Carilli, C.\ L., 
\& Ivison, R.\ J. 1999, \mnras, 305, 492}
 \reference{1}
{Eckart, A., \& Genzel, R. 1997, \mnras, 284, 576}
 \reference{1}
{Fabian, A.\ C., Blandford, R.\ D., Guilbert, P.\ W., 
Phinney, E.\ S., \& Cuellar, L. 1986, MNRAS, 221, 931}
 \reference{1} 
{Frank, J., King, A., \& Raine, D. 1992, Accretion Power 
in Astrophysics (Cambridge: Cambridge University Press)}
 \reference{1}
{Haller, J.\ W., Rieke, M.\ J., Rieke, G.\ H., Tamblyn, P., 
Close, L., \& Melia, F. 1996, \apj, 456, 194}
 \reference{1}
{Jones, F.\ C. 1968, Phys. Rev., 167, 1159}
 \reference{1}
{Kaburaki, O. 1999, in Disk Instabilities in Close Binary 
Systems, ed. S.\ Mineshige et al. (Tokyo: Universal Academy 
Press), 325 (K99)}
 \reference{1}
{Kaburaki, O. 2000, \apj, in press (astro-ph 9910252) (K00)}
 \reference{1} 
{Kaburaki, O., \& Itoh, M. 1987, A\&A, 172, 191}
 \reference{1}
{Kato, S., Fukue, J., \& Mineshige, S. 1998, Black-hole Accretion 
Disks (Kyoto: Kyoto University Press)}
 \reference{1}
{Krichbaum, T.\ P., Graham, D.\ A., Witzel, A., Greve, A., Wink, J.\ E., 
Grewing, M., Colomer, F., de Vicente, P., G\'{o}mez-Gonz\'{a}lez, J., 
Baudry, A., \& Zensus, J.\ A. 1998, A\&A, 335, L106}
 \reference{1}
{Lightman, A.\ P., \& Zdziarski, A.\ A. 1987, \apj, 319, 643}
 \reference{1}
{Lo, K.\ Y., Shen, Z.-Q., Zhao, J.-H., \& Ho, P.\ T.\ P. 1998, 
\apj, 508, L61}
 \reference{1}
{Mahadevan, R. 1997, ApJ, 477, 585}
 \reference{1}
{Mahadevan, R.  1999, \mnras, 304, 501 }
 \reference{1}
{Mahadevan, R., Narayan, R., \& Krolik, J.  1997, \apj, 486, 268}
 \reference{1}
{Mahadevan, R., Narayan, R., \& Yi, I. 1996, ApJ, 465, 327}
 \reference{1}
{Manmoto, T., Mineshige, S., \& Kusunose, M. 1997, ApJ, 489, 791}
 \reference{1}
{Nakamura, K.\ E., Kusunose, M., Matsumoto, R., \& Kato, S. 1997, 
\pasj, 49, 503}
 \reference{1}
{Narayan, R., \& Yi, I. 1994, ApJ, 428, L13}
 \reference{1}
{Narayan, R., \& Yi, I. 1995a, ApJ, 444, 231}
 \reference{1}
{Narayan, R., \& Yi, I. 1995b, ApJ, 452, 710}
 \reference{1}
{Narayan, R., Mahadevan, R., \& Quataert, E. 1998, in Theory of 
Black Hole Accretion Disks, ed. M.\ A.\ Abramowicz, G.\ Bjornsson, 
\& J.\ E.\ Pringle (Cambridge: Cambridge University Press), 148 
(astro-ph 9803141)}
 \reference{1}
{Narayan, R., Yi, I., \& Mahadevan, R. 1995, Nature, 374, 623}
 \reference{1}
{Narayan, R., Mahadevan, R., Grindlay, J.\ E., Popham, 
P.\ G.\ \& Gammie, C. 1998, \apj, 492, 554}
 \reference{1}
{Quataert, E., \& Narayan, R.  1999, \apj, 520, 298}
 \reference{1}
{Rybicki, G.\ B., \& Lightman, A.\ P. 1979, Radiative Processes 
in Astrophysics (New York: John Wiley \& Sons)} 
 \reference{1}
{Yusef-Zadeh, F., Morris, M., \& Chance, D. 1984, \nat, 310, 557}
 
\end{references}
\end{document}